\documentclass[sn-aps,iicol]{sn-jnl}

\usepackage{graphicx}%
\usepackage{multirow}%
\usepackage{amsmath,amssymb,amsfonts}%
\usepackage{amsthm}%
\usepackage{mathrsfs}%
\usepackage[title]{appendix}%
\usepackage{xcolor}%
\usepackage{textcomp}%
\usepackage{manyfoot}%
\usepackage{booktabs}%
\usepackage{listings}%

\begin{document}

\title[Estimation of collision centrality in HIC using deep learning]{Estimation of collision centrality in terms of the number of participating nucleons in heavy-ion collisions using deep learning}


\author[1,2]{\fnm{Dipankar} \sur{Basak}}\email{dipankar0001@gmail.com}

\author*[1]{\fnm{Kalyan} \sur{Dey}}\email{kalyn.dey@gmail.com}

\affil*[1]{\orgdiv{Department of Physics}, \orgname{Bodoland University}, \orgaddress{ \city{Kokrajhar}, \postcode{783370}, \state{Assam}, \country{India}}}

\affil[2]{\orgdiv{Department of Physics}, \orgname{Kokrajhar Govt. College}, \orgaddress{\city{Kokrajhar}, \postcode{783370}, \state{Assam}, \country{India}}}

\abstract{The abstract serves both as a general introduction to the topic and as a brief, non-technical summary of the main results and their implications. Authors are advised to check the author instructions for the journal they are submitting to for word limits and if structural elements like subheadings, citations, or equations are permitted.}


\abstract{The deep learning technique has been applied for the first time to investigate the possibility of centrality determination in terms of the number of participants ($N_{\mathrm{part}}$) in high-energy heavy-ion collisions. For this purpose, supervised learning using both deep neural network (DNN) and convolutional neural network (CNN) is performed with labeled data obtained by modeling relativistic heavy-ion collisions utilizing A Multi-phase Transport Model (AMPT). Event-by-event distributions of pseudorapidity and azimuthal angle of charged hadrons weighted by their transverse momentum are used as input to train the DL models. The DL models did remarkably well in predicting $N_{\mathrm{part}}$ values with CNN slightly outperforming the DNN model. The Mean Squared Logarithmic Error (MSLE) for the CNN model (Model-4) is determined to be 0.0592 for minimum bias collisions and 0.0114 for 0-60\% centrality class, indicating that the model performs better for semi-central and central collisions. Furthermore, the studied DL model is proven to be robust to changes in energy as well as model parameters of the input. The current study demonstrates that the data-driven technique has a distinct potential for determining centrality in terms of the number of participants in high-energy heavy-ion collision experiments.}




\maketitle

\section{Introduction}\label{sec1}
Investigations at various relativistic heavy-ion collision facilities suggest the formation of an extremely hot and dense state of strongly interacting matter where the quarks and gluons are no longer confined \cite{Xu:2014}. This short-lived state of matter known as Quark Gluon Plasma (QGP) is predicted to have existed barely a few microseconds after the Big Bang \cite{Chen:2014}. The favorable condition for the creation of such a hot and dense medium formed in heavy-ion collisions depends on the produced initial energy density. If the resulting energy density surpasses 1 $\mathrm{GeV}/\mathrm{fm^3}$, lattice QCD predicts a phase transition from hadronic matter to deconfined QGP \cite{Karsch:2002}. The production of such a hot and dense matter is only expected in head-on collisions between heavy ions at relativistic energies. In heavy-ion physics terminology, centrality is a quantity that describes the initial collision geometry and measures the amount of overlap between the colliding nuclei. Determining centrality, therefore, becomes critical for selecting relevant events in nucleus-nucleus collisions. Geometrically, centrality can be represented by a quantity called the impact parameter \footnote{Impact parameter is defined as the distance between centers of the two colliding nuclei in the plane transverse to the collision axis}. The total number of nucleons involved in a collision, represented by $N_{\mathrm{part}}$, is another proxy for centrality in heavy-ion collisions. Unfortunately, none of the aforementioned quantities are experimentally measurable due to the femtoscopic length scale of the system. These quantities can only be determined indirectly using the Glauber model which uses experimentally measured charge particle multiplicity or the forward energy carried by the spectator nucleons as the input \cite{Miller:2007}. In the following contribution, an alternative approach based on Deep Learning has been implemented for the first time to determine collision centrality in terms of the number of participating nucleons. 

The past several years have seen a rise in the use of machine learning (ML) and deep learning (DL) approaches to solve problems in a variety of modern science domains, including nuclear and high-energy physics \cite{Boehnlein:2022, Du:2020, Kuttan:2021, Zhao:2022, arxiv.2111.15655, Monk:2018, Shokr:2022, Pang:2016, Tsang:2021, Mallick:2021}. DL, a subset of AI, is a very powerful data science technique capable of identifying and learning essential hidden characteristics or patterns from complex nonlinear systems with high-order correlations \cite{Pang:2016}. The ML and DL models can learn pertinent attributes from data of their own without instructions being explicitly programmed. Data-driven techniques have become a useful tool for analyzing p-p, p-A, and heavy-ion physics data in parallel to conventional methods. ML and DL methods were successfully applied in the field of heavy-ion physics to determine various quantities of interest such as the impact parameter \cite{Bass:1996, Li:2020, Li:2021, Xiang:2022, Zhang:2022, Saha:2022, Kuttan:2021}, elliptic flow coefficient \cite{Mallick:2022, Mallick:2023}, transverse spherocity \cite{Mallick:2021}, the temperature of the system \cite{Song:2021} etc.

In 1996, Bass \textit{et al.} \cite{Bass:1996} demonstrated for the first time in their study that the prediction of impact parameters using neural network approaches is considerably superior to classical methods. Among the recent works, Li \textit{et al.} \cite{Li:2020, Li:2021} used  ANN (Artificial Neural Network), CNN, and LightGBM (gradient-boosting machine) algorithms \cite{Ke:2017} to estimate impact parameter from simulated  Au~+~Au collisions at intermediate energies (0.2-1.0 GeV/nucleon) and Sn+Sn collisions at the beam energy of 270 MeV/nucleon with UrQMD model \cite{Bass:1998, Bleicher:1999}. The accuracy of determination of impact parameter in their study was found to be less for central collisions. Other works include the calculation of impact parameter with the help of DNN and CNN algorithms utilizing AMPT \cite{Lin:2005} generated Au+Au collisions at $\sqrt{s}=200$~GeV  \cite{Xiang:2022}. Although good accuracy is observed in predicting the value of impact parameter, their model fails to achieve good accuracy for central collisions. Zhang \textit{et al.} on the other hand using the CNN model showed reasonable accuracy in determining impact parameter throughout the entire centrality range at low-intermediate beam energies \cite{Zhang:2022}. ML technique such as the Boosted Decision Trees (BDTs) \cite{Friedman:2001} has been implemented by Mallick \textit{et al.} \cite{Mallick:2021} to predict impact parameter and transverse spherocity in Pb+Pb collisions at the LHC energies. Their model could able to predict both spherocity and impact parameter with Mean Absolute Error values of 0.055 and 0.52 fm respectively. On the other hand, Saha \textit{et al.} \cite{Saha:2022} used three different standard ML algorithms namely k-nearest-neighbors (kNN), extra-trees regressor (ET), and the random forest regressor (RF) model to determine impact parameter, eccentricity, and participant eccentricity at LHC energies. Their method is found to be more accurate in determining impact parameter than the standard application of ML training methods. Another variant of deep learning models known as PointNet model \cite{point_net:2016} was applied at FAIR energies (10\textit{A} GeV) to determine impact parameter on an event-by-event basis for the proposed CBM experiment \cite{Ablyazimov:2017, Senger:2020, claudia:2007}. The model was trained on features like the hit position of particles in the CBM detector planes, tracks reconstructed from the hits etc., and found to be more accurate and less model dependent than conventional methods \cite{Kuttan:2021}.
 
In this present work, an attempt has been made to estimate the number of participating nucleons ($N_{\mathrm{part}}$) in high-energy heavy-ion collisions with the help of a deep learning technique. The paper is structured as follows. In section \ref{sec2}, we discuss the methodology of the present investigation that includes event generation, a brief introduction to the AMPT model, and the input selection criteria for training the studied DL models. We also presented a quick overview of the general architecture of DL models in section \ref{sec3}.  In section \ref{sec4}, we presented the results where we also discussed the performance of the studied DL models with centrality, energy, input grid dimension, and on event-generator configuration. Finally, in section \ref{sec5}, we summarize our findings.

\section{Methodology}\label{sec2}
\subsection{Event Generation}\label{subsec1}
A Multi-Phase Transport Model (AMPT) \cite{Lin:2005} was employed for simulating minimum bias heavy-ion collisions used as inputs for the present investigation. AMPT is a general-purpose Monte Carlo transport model used to simulate p+p, p+A, and heavy-ion collisions at relativistic energies. The model consists of four major components: initial conditions, partonic interactions, hadronization, and hadronic interactions. The initial conditions are generated according to Heavy Ion Jet Interaction Generator (HIJING) model \cite{Wang:1991}, which includes  the initial spatial and momentum distributions of hard minijet partons and soft string excitations. The interactions among partons are described by Zhang’s parton cascade (ZPC) model \cite{Zhang:1998}. In the default version of the AMPT model, the hadronization process is implemented via the Lund string fragmentation scheme \cite{Andersson:1983} and in the string melting version (AMPT-SM), it is done by the quark coalescence model \cite{He:2017}. Finally, the hadronic interactions are modeled via a relativistic transport model (ART) \cite{Li:1995}. For the present study, both the versions of the AMPT model i.e. default and string melting configuration  were used where minimum bias Au+Au collisions at different beam energies were generated.

\begin{figure}[tb]
\includegraphics[width=0.44\textwidth]{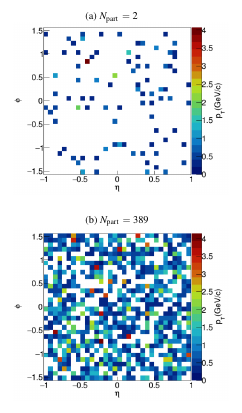}
\caption{(Color online) $p_T$ weighted $\left(\eta - \phi \right)$ spectra for extreme peripheral (upper panel) with $N_{\mathrm{part}}=2$ and for most central (lower panel) events with $N_{\mathrm{part}}=389$ using AMPT-SM in  Au+Au collisions at $\sqrt{s}=200$ GeV. The spectra shown here are with $32 \times 32$ grid  dimensions.}. \label{fig1}
\end{figure}

\subsection{Input Selection}\label{subsec2}
The selection of appropriate input is critical for any DL regression task. The input variables must be chosen so that they are highly correlated with the target variable. Different kinematical parameters of the produced particles, such as transverse momentum ($p_T$), pseudorapidity ($\eta$), and azimuthal angle ($\phi$) are expected to retain information about the initial conditions of heavy-ion collisions. Hence, for the present investigation, the phase space distributions of the copiously produced charged hadrons ($\pi^\pm$, $k^\pm$, $p$, $\Bar{p}$) were considered as input to the DL models. We have used $p_T$ weighted $\left(\eta - \phi \right)$ spectra on event-by-event basis. To mimic the realistic experimental conditions, charged hadrons within mid-rapidity ($\lvert \eta \rvert <1$) acceptance which corresponds to the STAR TPC detector  acceptance \cite{Anderson:2003} and a transverse momentum cut of ($p_T>0.2$ GeV/c) were chosen. In order to show two contrasting scenarios, $p_T$ weighted $\left(\eta - \phi \right)$ spectra with $32 \times 32$ grid  dimensions for both extreme central (lower panel) and most peripheral (upper panel) collisions are shown in Fig.~\ref{fig1}.

\section{Deep Learning Models}\label{sec3}
In this section, we have presented a general introduction to the DL models and discussed explicitly the architectures of different models used in the present study. Two of the most efficient DL models namely Deep Neural Networks (DNN) and Convolutional Neural Networks (CNN) are used to estimate $N_{\mathrm{part}}$ in high-energy heavy-ion collisions. DNN or a multilayer perceptron (MLP) is one of the simplest versions of DL models which finds application in diverse fields such as machine translation, speech recognition, and computer vision problems, just to name a few. It comprises an input layer, several hidden layers or fully-connected layers (FC), and an output layer made up of neurons or nodes. The inputs to each one of the neurons in a layer are the outputs of all of the neurons from the preceding layer. On the other hand, CNN is another powerful class of neural networks specially used for spatial pattern recognition problems. A CNN consists of several layers where the output of each layer serves as the input for the next layer. Three types of layers are generally used in a CNN model namely convolution layers, pooling layers, and fully-connected layers. The convolution layer includes kernels/filters that detect features from the input image and store the results in a so-called feature map or convolved feature. The last few layers of a CNN consist of dense layers or fully connected layers. A fully connected layer takes the inputs from the convolution layer and applies weights to predict the correct label.  In addition to that, in CNNs, a pooling layer \cite{Scherer:2010} is generally introduced after the convolution layers to decrease the spatial dimension of the convolution output and hence to reduce the computational complexity. Activation functions are always added with each layer to introduce non-linearity into the output of each layer. Sigmoid, tanh, and ReLU or rectified linear unit etc. are some of the commonly used activation functions. In supervised learning, the difference between the predicted output of the network and the true value is evaluated using a loss function. During the training of the neural network, the loss is minimized using an optimizer algorithm that updates the values of the parameters at every stage of the training. 

The learning outcome of a DL model depends on the choice of the architecture of the model i.e. the combinations of various hyperparameters such as the number of layers, number of pooling layers, number of fully-connected layers, the number of neurons per layer, dropout rate, learning rate, type of optimizer, etc. We checked different models with different combinations of hyperparameters values and complexity. Based on the learning performance, four DL models were finally selected for the current supervised regression task. The architecture of each of the models used in the current work is depicted in Table~\ref{tab1}.
 
In all four models, for the input and hidden layers, ReLU \cite{Nair:2010} is used as the activation function  while for the output layer linear activation function is considered. In all the CNN models, Average-pooling of pool size $2 \times 2$ was used as pooling layers. The most efficient Adam optimization algorithm \cite{Kingma:2015} with a learning rate of 0.0001 is employed for the training of the networks and Mean-Squared Error (MSE) is chosen as the loss function. The models were allowed to train with a batch size of 64. As over-fitting is one of the common problems of any neural network, to overcome this problem we have used (i) L1 regularization \cite{Tibshirani:1996} and L2 regularization \cite{Hinton:1986}, (ii) Batch normalization (BN) \cite{Ioffe:2015}, (iii) Dropout layers \cite{Hinton:2014} and (iv) EarlyStopping mechanism \cite{Yao:2007} with patience level 10. The L1 and L2 regularization prevent the weights and biases from increasing to arbitrarily large values during the optimization whereas  Batch normalization normalizes the output of a layer with additional scaling and shifting, thereby increasing the rate of learning  of the model by reducing the number of epochs required. EarlyStopping callback, on the other hand, is used to terminate the training process as soon as the validation error reaches its minimum value. Dropout is another popular regularization technique where some number of nodes in the layers are randomly ignored during the training process. Training of the models was performed using Keras (Version: 2.9.0) \cite{Chollet} with Tensorflow backend (Version: 2.9.1) \cite{Abadi:2016}.

\begin{table*}[h!]
\begin{minipage}{\textwidth}
\caption{Architecture of all the DL models used in the present investigation.}\label{tab1}
\begin{tabular*}{\textwidth}{@{\extracolsep{\fill}}ccccc@{\extracolsep{\fill}}} 
\toprule%
Models  & Model Type    & Layers            & Description          & Total Parameters\footnote{Trainable + Non-trainable parameters.} \\
\midrule
Model-1 & DNN           & Input             & 1 × 3072 tensor        & 1,737,729 \\
        &               & FC, ReLU          & 512 neurons \\
        &               & FC, ReLU          & 256 neurons \\
        &               & FC, ReLU          & 128 neurons  \\
        &               & FC                & 1 output\\
\midrule
Model-2 & CNN           & Input             & 3 × 32 × 32 tensor     & 47,329 \\
        &               & Conv2D, ReLU, BN  & 32 kernels (3 × 3) \\
        &               & AveragePool2D     & 2 × 2 kernel, stride 2 \\
        &               & Dropout           &  0.2 \\
        &               & Conv2D, ReLU, BN  & 32 kernels (3 × 3) \\
        &               & AveragePool2D     & 2 × 2 kernel, stride 2 \\
        &               & Dropout           &  0.2 \\
        &               & Flatten           & 1152 neurons \\
        &               & FC, ReLU          & 32 neurons \\
        &               & FC                & 1 output \\
\midrule
Model-3 & CNN           & Input                 & 3 × 30 × 30 tensor    & 623,201 \\
        &               & Conv2D, ReLU, BN      & 16 kernels (3 × 3) \\
        &               & Conv2D, ReLU, BN      & 32 kernels (3 × 3) \\
        &               & AveragePool2D         & 2 × 2 kernel, stride 2 \\
        &               & Dropout               &  0.2 \\
        &               & Conv2D, ReLU, BN      & 64 kernels (3 × 3) \\
        &               & AveragePool2D         & 2 × 2 kernel, stride 2 \\
        &               & Dropout               &  0.2 \\
        &               & Conv2D, ReLU, BN      & 128 kernels (3 × 3) \\
        &               & Dropout               &  0.1 \\
        &               & Flatten               & 2048 neurons \\
        &               & FC, ReLU              & 256 neurons \\
        &               & FC                    & 1 output \\
\midrule
Model-4 & CNN           & Input  & 3 × 30 × 30 tensor        & 1,274,305 \\
        &               & Conv2D, ReLU , BN & 32 kernels (3 × 3) \\
        &               & Dropout           &  0.2 \\
        &               & Conv2D, ReLU, BN  & 64 kernels (3 × 3) \\
        &               & AveragePool2D     & 2 × 2 kernel, stride 2 \\
        &               & Dropout           &  0.2 \\
        &               & Conv2D, ReLU, BN  & 128 kernels (3 × 3)  \\
        &               & AveragePool2D     & 2 × 2 kernel, stride 2 \\
        &               & Dropout           &  0.1 \\
        &               & Flatten           & 4608 neurons \\
        &               & FC, ReLU          & 256 neurons \\
        &               & FC                & 1 output  \\
\end{tabular*}
\end{minipage}
\end{table*}

\begin{figure*}[tb]
\includegraphics[width=0.9\textwidth]{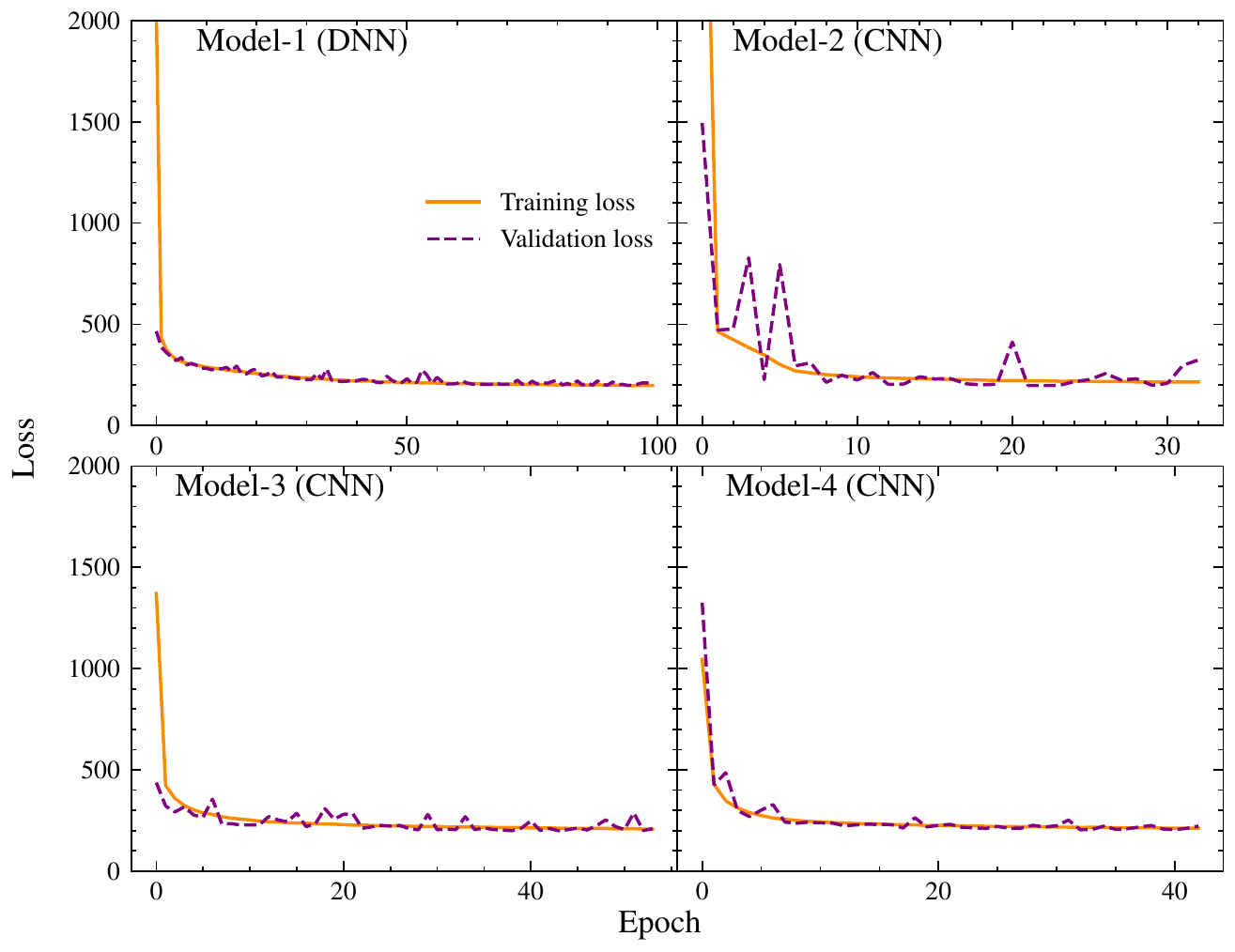}
\caption{(Color online) Model loss in terms of mean squared error as a function of number of epochs for all the DL models used (see text for details).}\label{fig2}
\end{figure*}

\begin{figure*}[h!]%
\centering
\includegraphics[width=0.9\textwidth]{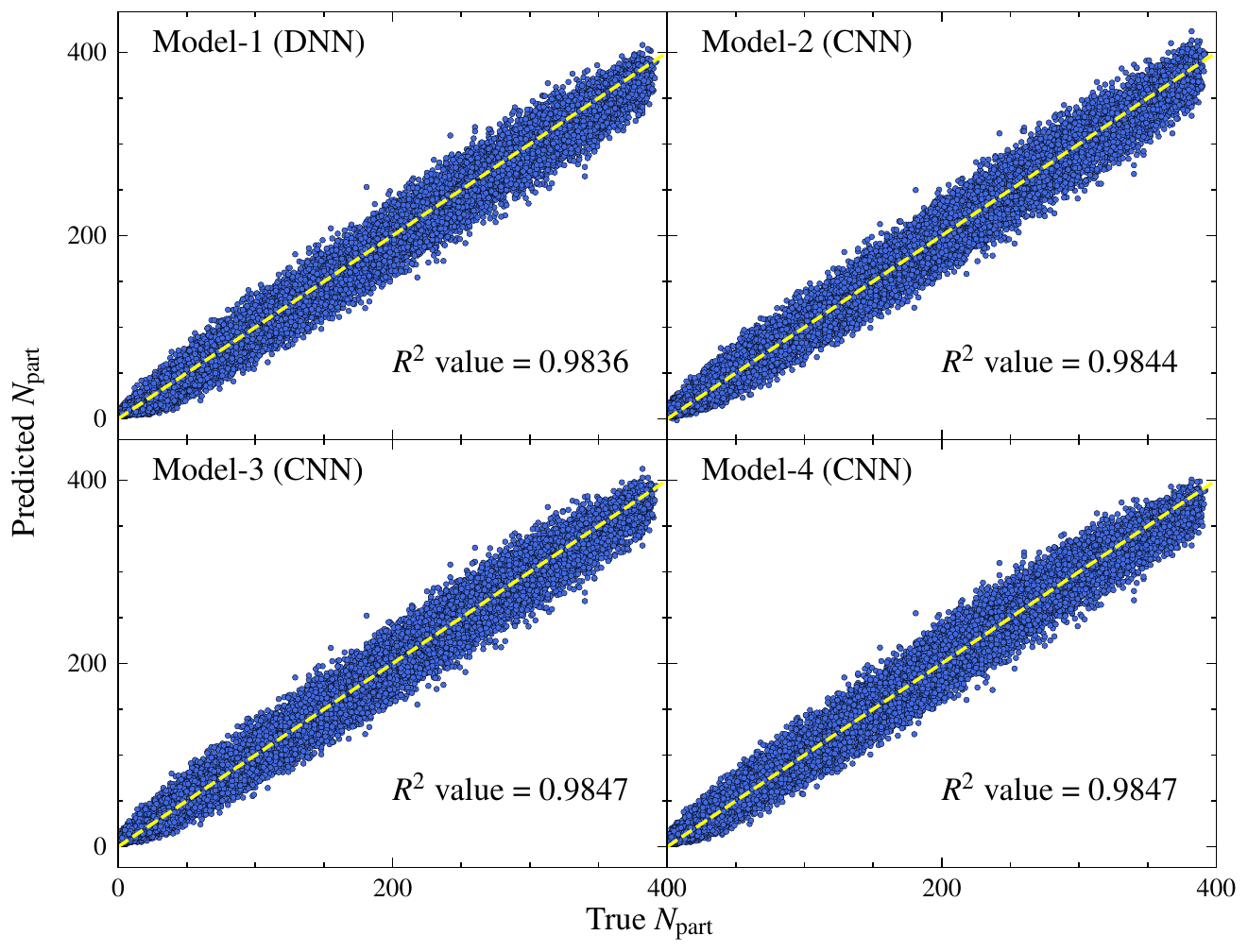}
\caption{(Color online) Correlation between the true and predicted $N_{\mathrm{part}}$ for all the DL models considered in the present study. The $R^2$ values are also shown.}\label{fig3}
\end{figure*}

\section{Results and Discussions}\label{sec4}
To begin with, 180K Au+Au collisions at $\sqrt{s}=200$ GeV were generated using AMPT-SM configuration. Out of which 30K events were kept aside for testing alone. The remaining 150K generated events were randomly split into training and validation data-set in the ratio of 3:1. The training and validation learning curves for all four DL models are shown in Fig.~\ref{fig2}. It is evident from the figure that for all four models considered in this study, the training and validation loss decreases sharply with the number of epochs. This implied that the studied DL models are quite successful in learning hidden features from the input training data set. It is further observed from this figure that after traversing through some epochs, both the training and validation losses match roughly with each other. This indicates that the trained models generalize well in predicting the target variable using the validation data-set. The next immediate task is to check the goodness of fitting of the simulated data with the studied linear regression models. This is done by calculating $R^2$-values also known as the coefficient of determination which is defined as,

\begin{equation}
R^2 = 1 - \frac{\sum_{i=1}^{N}{\left(y_i^t-y_i^p\right)}^2}{\sum_{i=1}^{N}{\left(y_i^t- \langle y^t \rangle \right)}^2}.
\label{eq1}
\end{equation}

Here $y$ and $N$ refer to the target variable and the number of events in  the test data-set respectively.  $y_i^t$ and $y_i^p$ represent respectively the true and predicted values of the target variable. The angular bracket $\langle \quad \rangle$ in Eq.~\ref{eq1} represents the mean over all the test data samples. On average, the $R^2$ value is observed to be 0.98 for the studied DNN and CNN models which indicates a good fitting of the simulated data with the linear regression models. But $R^2$ measure alone may not determine if a regression model gives an adequate fit to the simulated data. $R^2$ measure in addition to the other statistical metric can provide a complete picture. The performance of the studied models has therefore been quantified using Mean Squared Logarithmic Error (MSLE) as depicted in Table~\ref{tab2}. The expression for MSLE metric is given by,
\begin{equation}
\text{MSLE} = \frac{1}{N}\sum_{i=1}^{N} {\{ \log \left(1+y_i^{t}\right)-\log\left (1+y_i^{p}\right)\}}^2.
\label{eq2}
\end{equation}
We use MSLE as it emphasizes the relative differences between predicted and actual values, instead of their absolute differences. This property is especially useful when the magnitude of the target variable varies significantly across the data-set.

\begin{table}[h!]
\caption{Performance of the DL models for 150~K training events and 30~K test events.}\label{tab2}%
\begin{tabular}{@{}lllc@{}}
\toprule
Models.     & MSLE      & $R^2$         & Average Time/epoch \\
\midrule
Model-1     & 0.0599    & 0.9836        & 28 sec\\
Model-2     & 0.0698    & 0.9844        & 43 sec\\
Model-3     & 0.0629    & 0.9847        & 85 sec\\
Model-4     & 0.0592    & 0.9847        & 188 sec \\
\botrule
\end{tabular}
\end{table}

It is seen from Table~\ref{tab2}, that all the DL models considered in the study have performed reasonably well in predicting $N_{\mathrm{part}}$ values with MSLE of about 0.06. It can further be noted that Model-4 (CNN) performs marginally better than the other models although it took a comparatively longer CPU time for execution. The large computational time taken may be due to the complexity of the model (with around 1.2 million model parameters). It may be noted that the present DL models were not optimized for speed. The study has been performed with the help of a CPU with Intel(R) Core(TM) i5-10300H processor and 16 GB DDR4 RAM.

Fig.~\ref{fig3} shows the correlation plots between the true and the predicted values of the $N_{\mathrm{part}}$ for the DL models estimated from 30K minimum bias Au+Au collisions using AMPT-SM. The yellow dashed line corresponds to ideal prediction i.e. $N_{\mathrm{part}}^{true}=N_{\mathrm{part}}^{pred.}$. As expected, the predicted values of $N_{\mathrm{part}}$ for all the DL models are well populated around the dashed lines suggesting reasonably good prediction of $N_{\mathrm{part}}$ by the studied DL models.

\begin{figure*}[h!]%
\centering
\includegraphics[width=0.98\textwidth]{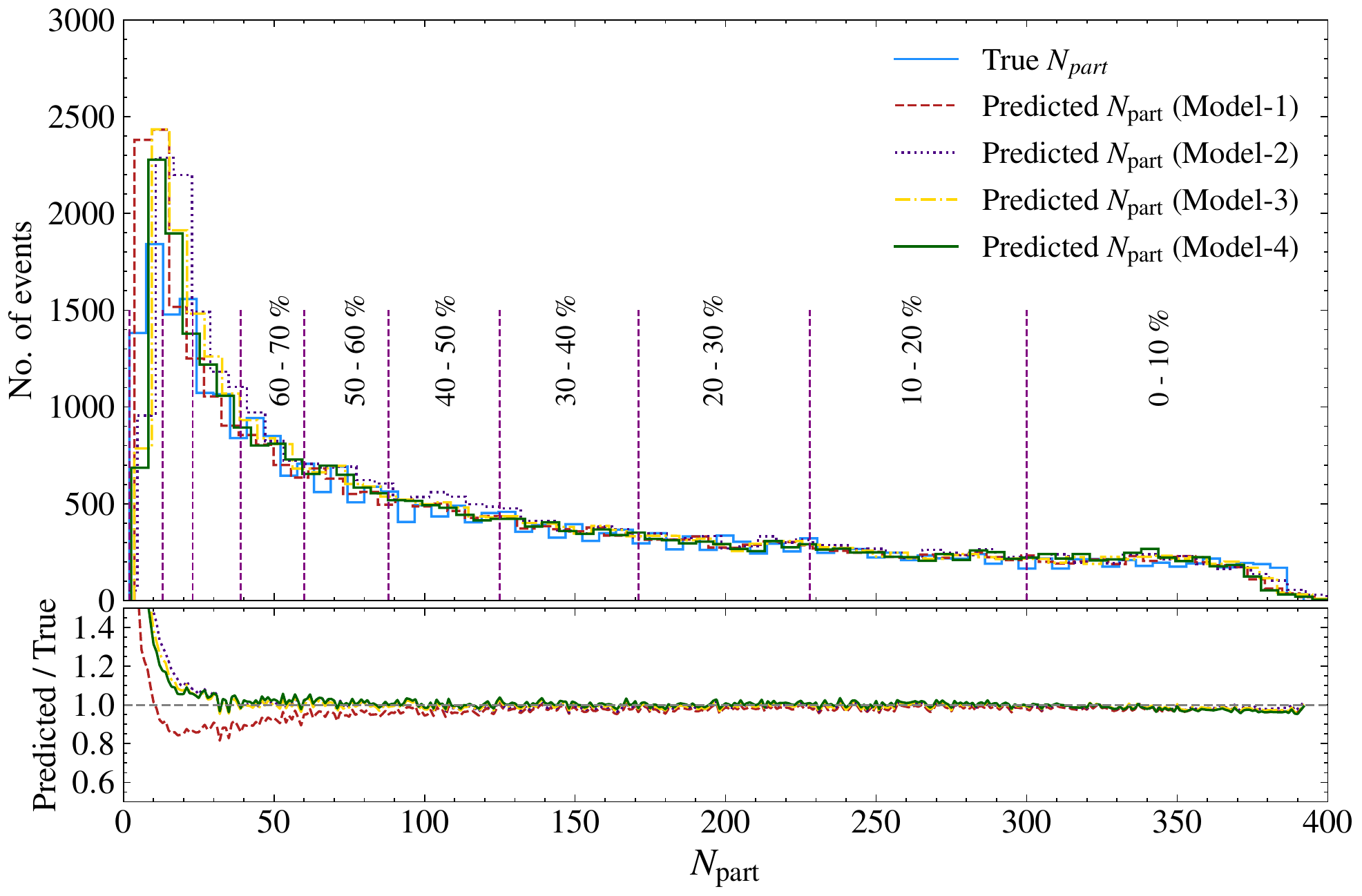}
\caption{(Color online) A comparison of true and predicted $N_{part}$ distribution for the DL models. The vertical slices represent the centrality percentiles.  The panel on the bottom shows the ratio of the predicted and actual $N_{part}$ as a function of true $N_{part}$.}\label{fig4}
\end{figure*}

\subsection{Centrality dependence}
In this section, we discuss the accuracy of the DL models at different collision centralities. In order to do that the predicted $N_{\mathrm{part}}$ distribution is plotted for all the models and compared with the ground truth as shown in the upper panel Fig.~\ref{fig4}. For better understanding, the ratio of predicted and true values of $N_{\mathrm{part}}$ is plotted against its true value as depicted in the lower panel of Fig.~\ref{fig4}. It is seen from this figure that all the DL models perform relatively better in semi-central and central collisions while the accuracy reduces for peripheral collisions. Further, all three CNN models perform better for wider centrality range $(0-60\%)$ while the DNN can provide a reasonably accurate estimation of the $N_{\mathrm{part}}$ in a comparatively narrow centrality window $(0-40\%)$.

\begin{figure}[tb]
\centering
\includegraphics[width=0.5\textwidth]{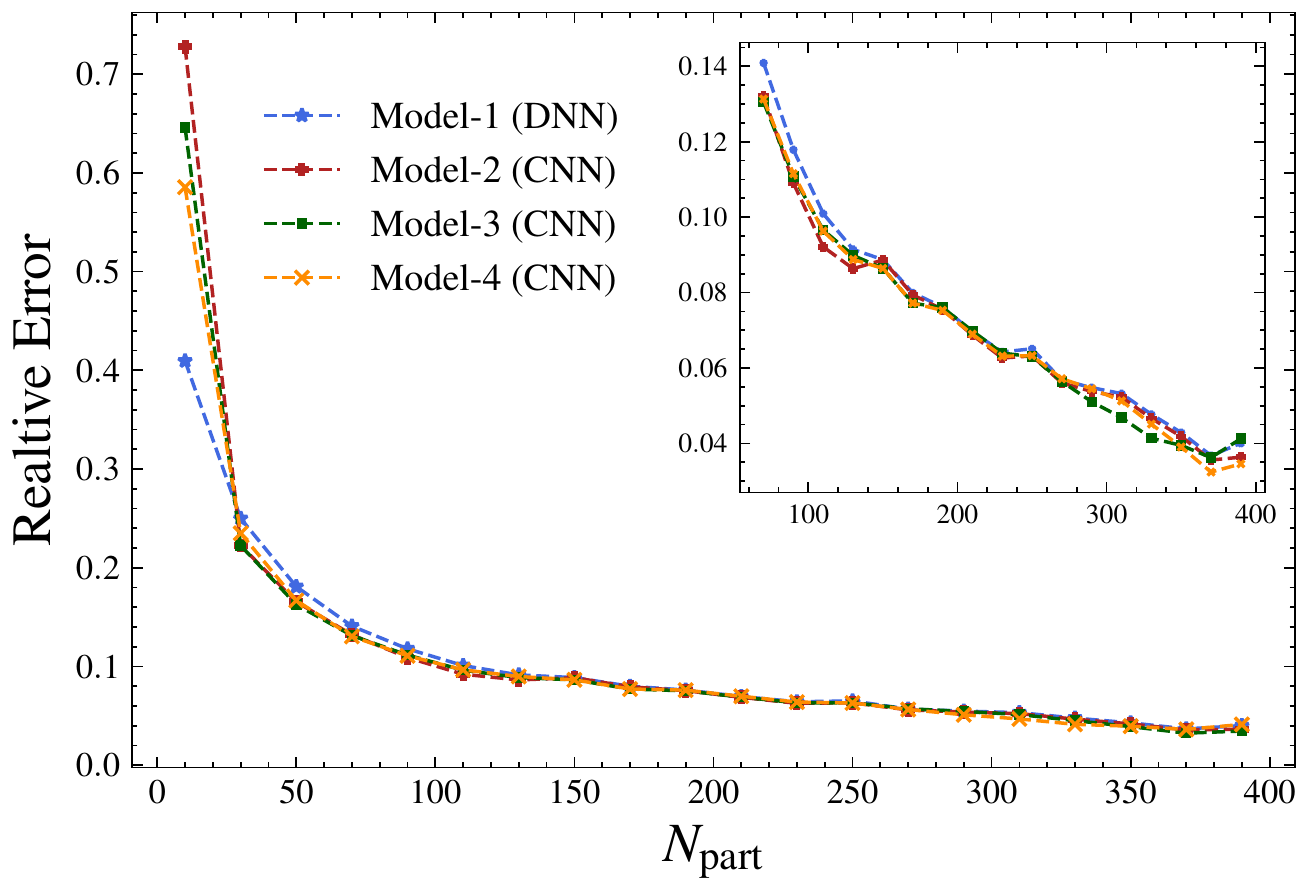}
\caption{(Color online) The variation of relative error against the actual value of $N_{\mathrm{part}}$ using the DL models. An inset plot is drawn for better visibility which magnifies the centrality window (0-60\%).}\label{fig5}
\end{figure}

In order to quantify the accuracy of the DL models, standard metrics such as relative error and MSLE (Eq.~\ref{eq2}) are calculated and tabulated in Table~\ref{tab3}. The relative error is defined as,
\begin{align}
  \text{Relative error} = \frac{\mid N_{\mathrm{part}}^t - N_{\mathrm{part}}^p  \mid}{N_{\mathrm{part}}^t}.                      
\end{align}
where $N_{\mathrm{part}}^t$ and $N_{\mathrm{part}}^p$ are the true and predicted value of $N_{\mathrm{part}}$ respectively. The relative error in $N_{\mathrm{part}}$ is plotted as a function of its actual values for all the studied DL models as shown in Fig.~\ref{fig5}. The relative error is seen to decrease sharply with the increase of $N_{\mathrm{part}}$ for all the DL models. From table~\ref{tab3}, it can be seen that for all the models, the MSLE value and mean relative error (MRE) for the centrality range 0 to 60\% are approximately 0.01 and 0.08 respectively while the MSLE and MRE values are higher for minimum bias collisions. The studied DL models, therefore, yield better results for semi-central and central collisions.

\begin{table}[h]
\caption{Centrality dependence of the performance of the models.}\label{tab3}%
\begin{tabular}{@{\extracolsep\fill}lclcc@{\extracolsep\fill}}
\toprule%
& \multicolumn{2}{c}{Minimum bias} & \multicolumn{2}{c}{0-60\% centrality} \\
\cmidrule{2-3} \cmidrule{4-5}%
Models.     & MSLE      & MRE        & MSLE           & MRE\\
\midrule
Model-1     & 0.0599    & 0.1716    & 0.0141        & 0.0846\\
Model-2     & 0.0698    & 0.2202    & 0.0117        & 0.0809\\
Model-3     & 0.0629    & 0.2055    & 0.0118        & 0.0810\\
Model-4     & 0.0592    & 0.1970    & 0.0114        & 0.0808\\
\botrule
\end{tabular}
\end{table}

\subsection{Input grid dimension dependence}
The performance of the DL model is also examined by varying the grid dimensions of the input spectra used for training. It can be noted from the Table~\ref{tab4}, that the performance of the DL model (Model-4) increases with increasing grid dimension and remains virtually constant for higher grid dimensions such as $32 \times 32$, $64 \times 64$ and $80 \times 80$. It is also seen from Table~\ref{tab4}, that the computational time increases exponentially with the increase of input grid dimension. Considering both factors, the entire study is being performed with grid dimensions of $32 \times 32$.

\begin{table}[h]
\caption{Performance of the CNN model (Model-4) for different grid dimensions of the inputs with $\eta-\phi$ spectra for 50K training events and 10K test events.}\label{tab4}%
\begin{tabular}{@{}lllc@{}}
\toprule
Grid dimension   & $R^2$    & MSLE          & Average Time/epoch \\
\midrule
$16\times16$    & 0.9767    & 0.0700        & 12 sec  \\
$32\times32$    & 0.9847    & 0.0579        & 68 sec  \\
$64\times64$    & 0.9852    & 0.0557        & 311 sec \\
$80\times80$    & 0.9840    & 0.0591        & 533 sec \\
\botrule
\end{tabular}   
\end{table}

\begin{table}[h!]
\caption{Performance of the CNN model (Model-4) for different beam energies $\sqrt{s}=7.7, 39, 64.2, 130, 200$ GeV.}\label{tab5}%
\begin{tabular}{@{}lll@{}}
\toprule
Energy (GeV)         & $R^2$    & MSLE           \\
\midrule
7.7                 & 0.9668    & 0.1790          \\
39                  & 0.9813    & 0.0875           \\
62.4                & 0.9815    & 0.0878           \\
130                 & 0.9829    & 0.0730           \\
200                 & 0.9842    & 0.0625         \\
\botrule
\end{tabular}
\end{table}

\begin{figure}[b!]%
\centering
\includegraphics[width=0.49\textwidth]{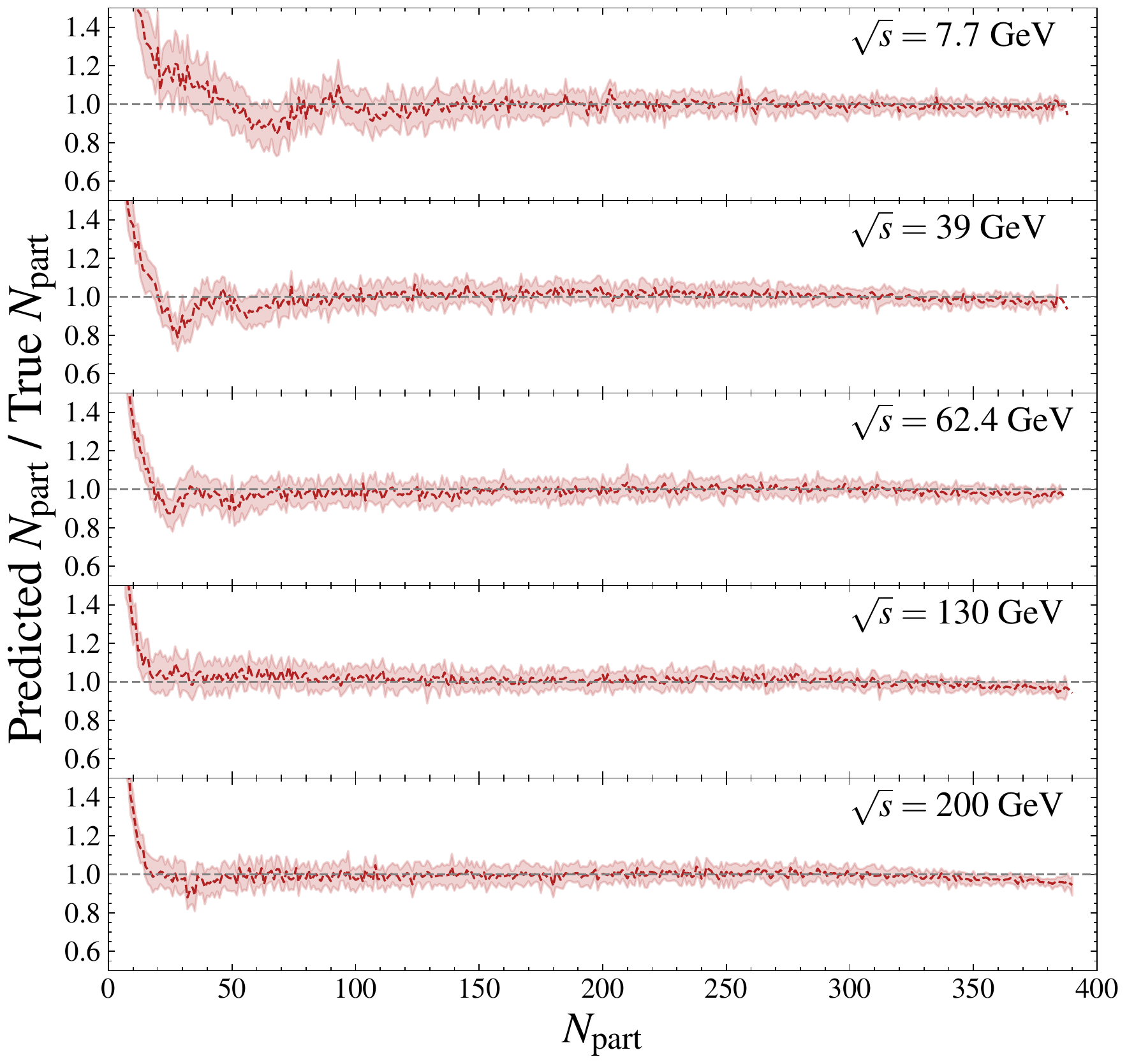}
\caption{(Color online)  The ratio of predicted and true $N_{part}$ as a function of true $N_{part}$ for various beam energies such as $\sqrt{s}=7.7, 39, 62.4, 130, 200$~ GeV.}\label{fig6}
\end{figure}

\subsection{Energy dependence}
In this section, we discuss the performance of the DL model at different beam energies. The best-performed model i.e Model-4 is used for training and testing the simulated events generated with AMPT-SM at RHIC BES (Beam Energy Scan) energies i.e. $\sqrt{s}=7.7, \ 39, \ 62.4, \ 130$~ GeV. For this present comparison, the DL model at all energies is trained and tested with  80K and 20K events respectively. In order to quantify the model performance, the metrics such as $R^2$ and MSLE values are calculated and tabulated in Table~\ref{tab5}. Moreover, in Fig.~\ref{fig6}, we have plotted the ratio of predicted and true $N_{\mathrm{part}}$ as a function of the true $N_{\mathrm{part}}$ values for all the energies. The model performs quite well in predicting $N_{\mathrm{part}}$ values at all energies as is evident from Table~\ref{tab5} and Fig.~\ref{fig6}. A slight decrease in the performance is observed at the lowest beam energy i.e. $\sqrt{s}=7.7$~ GeV. Also, at all beam energies, the performance is better towards central collisions. The above two results can be explained from the fact that both at higher beam energies and at more central collisions more charged particles are produced thereby proving more information to the neural network to process for predicting $N_{\mathrm{part}}$ values.

\begin{figure}[b!]
\includegraphics{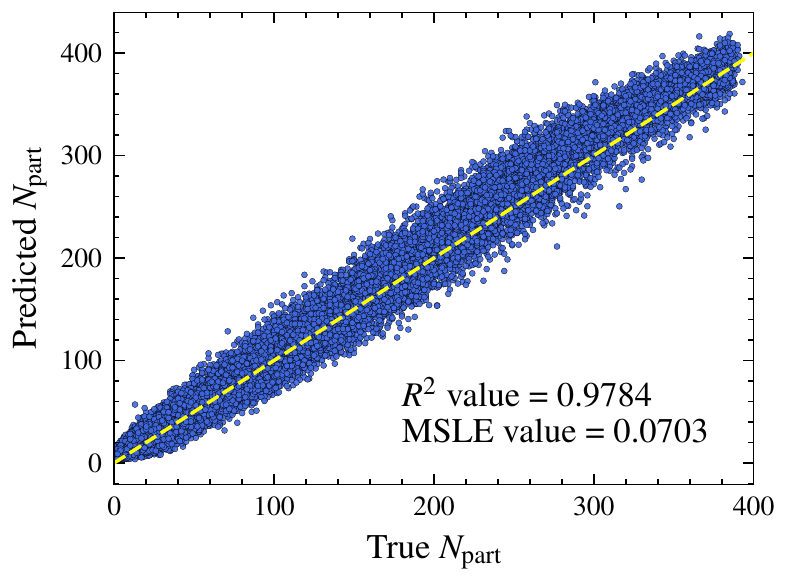}
\caption{(Color online) Correlation between the true and predicted $N_{\mathrm{part}}$ with Model-4 trained using AMPT-SM but tested with AMPT default events.}
\label{fig7}
\end{figure}

\subsection{Dependence on event-generator configuration}
To test the robustness of the DL model against changes in the event-generator configuration, a study was conducted where the CNN model (Model-4) trained with AMPT-SM was tested with events generated with AMPT default configuration. It is to be noted that although the initial collision configuration of both AMPT-SM and AMPT default are similar, both models evolved with different physics approaches. In the default configuration, only the minijet patrons take part in ZPC and eventually recombine with parent strings by Lund string fragmentation. On the other hand, in string melting configuration, all the excited strings from the HIJING are converted into partons which increases the parton density in ZPC and hadronization takes place via the quark coalescence model. Fig.~\ref{fig7} shows the correlation between true and predicted $N_{\mathrm{part}}$ with Model-4 trained using AMPT-SM but tested with AMPT default events. The $R^2$ and MSLE values of the prediction are found to be 0.9784 and 0.0703 respectively from which it is clear that in spite of the differences in physics perspective for both the AMPT-SM and AMPT default models, the present DL model could able to generalize well in predicting $N_{\mathrm{part}}$ values for the AMPT default configuration.

\section{Summary}\label{sec5}
The deep learning technique has been employed for the first time to determine centrality in terms of $N_{\mathrm{part}}$ in high-energy heavy-ion collisions. For this study, we have used two  variants of DL models such as DNN and CNN. 
$p_T$ weighted $\left(\eta - \phi \right)$ spectra of charged hadrons produced in minimum bias Au+Au collisions at $\sqrt{s}=200$ GeV generated with AMPT-SM model were used to train the DL models. Although the studied DL models did perform remarkably well in estimating $N_{\mathrm{part}}$, especially for semi-central and central collisions, the CNN provides marginally better results. To cross-check the performance of the DL model at other energies, the best-performing model (model-4) without altering its architecture was trained and tested at different RHIC-BES energies, namely $\sqrt{s}=7.7, 39, 64.2, 130$ GeV. Except for the lowest beam energy, the performance of the DL model at other energies is identical to that of the top RHIC energy i.e. $\sqrt{s}=200$ GeV as demonstrated by the $R^2$ and MSLE values. Further, the efficacy of the studied DL model is also investigated by changing the event-generator configuration. The DL model (model-4) trained using AMPT-SM is therefore tested with the AMPT default configuration and predicts $N_{\mathrm{part}}$ values with reasonable accuracy, indicating that the studied DL model is robust to changes in event-generator configuration.


\end{document}